\documentclass{article}

\usepackage[preprint]{neurips_2025}
\usepackage[utf8]{inputenc} % allow utf-8 input
\usepackage[T1]{fontenc}    % use 8-bit T1 fonts
\usepackage{hyperref}       % hyperlinks
\usepackage{url}            % simple URL typesetting
\usepackage{booktabs}       % professional-quality tables
\usepackage{amsfonts}       % blackboard math symbols
\usepackage{nicefrac}       % compact symbols for 1/2, etc.
\usepackage{microtype}      % microtypography
\usepackage{xcolor}         % colors

\usepackage{mathtools}
\usepackage{graphicx}
\usepackage{multirow}
\usepackage{wrapfig}

\usepackage{algorithm}%

\usepackage{url}            % simple URL typesetting
\usepackage{algorithmic}   % for \STATE and algorithmic environment

\usepackage{adjustbox}
\usepackage{subcaption}
\usepackage{enumitem}

\usepackage{amsmath}
\usepackage{amssymb}
\usepackage{nomencl}
\usepackage{glossaries}
\usepackage{xspace}
\usepackage{multirow}
\usepackage{graphicx}
\usepackage{bbding}
\usepackage{pifont}
\usepackage{enumitem}
\usepackage{multicol}
\usepackage{array}
\usepackage{fp}
\usepackage{makecell}
\usepackage{flushend}
\usepackage{cases}
\usepackage{color}
\usepackage{amsmath}
\usepackage{bm}      % 加粗数学符号
\usepackage{bbm}     % 用于 \mathbbm{1} 样式的指示函数
\usepackage{amsmath,amsfonts,bm}

\newcommand{\bs}{\mathbf{s}}
\newcommand{\ba}{\mathbf{a}}

\newcommand{\bx}{\mathbf{x}}
\newcommand{\bc}{\mathbf{c}}
\newcommand{\bp}{\mathbf{p}}

\definecolor{lightpink}{HTML}{FF99CC}
\definecolor{lightblue}{HTML}{A680B8}
\definecolor{cyans}{HTML}{5CD7E0}

\usepackage{amsmath,amsfonts,bm}

\def\eqref#1{equation~\ref{#1}}
\def\1{\bm{1}}

\DeclareMathAlphabet{\mathsfit}{\encodingdefault}{\sfdefault}{m}{sl}
\SetMathAlphabet{\mathsfit}{bold}{\encodingdefault}{\sfdefault}{bx}{n}

\newcommand{\R}{\mathbb{R}}

\title{ProtInvTree: Deliberate Protein Inverse Folding \\with Reward-guided Tree Search}

\vspace{-0.2cm}
\author{Mengdi Liu\textsuperscript{1,2}, Xiaoxue Cheng\textsuperscript{4}, Zhangyang Gao\textsuperscript{3}, Hong Chang\textsuperscript{1,2}\thanks{Corresponding author}~, \\
\textbf{Cheng Tan\textsuperscript{3},
Shiguang Shan\textsuperscript{1,2}, Xilin Chen\textsuperscript{1,2}} \\
\textsuperscript{1} Institute of Computing Technology, Chinese Academy of Sciences\\
\textsuperscript{2} University of Chinese Academy of Sciences\\
\textsuperscript{3} AI Lab, Research Center for Industries of the Future, Westlake University\\ 
\textsuperscript{4} Gaoling School of Artificial Intelligence, Renmin University of China\\ 
\texttt{\{liumengdi23z, changhong, sgshan, xlchen\}@ict.ac.cn}, \\
\texttt{chengxiaoxue@ruc.edu.cn},
\texttt{\{gaozhangyang, tancheng\}@westlake.edu.cn}\\
}

\begin{document}

\maketitle

\vspace{-0.3cm}
\begin{abstract}
    Designing protein sequences that fold into a target 3D structure—known as protein inverse folding—is a fundamental challenge in protein engineering. While recent deep learning methods have achieved impressive performance by recovering native sequences, they often overlook the one-to-many nature of the problem: multiple diverse sequences can fold into the same structure. This motivates the need for a generative model capable of designing diverse sequences while preserving structural consistency. To address this trade-off, we introduce \textbf{ProtInvTree}, the first reward-guided tree-search framework for protein inverse folding. ProtInvTree reformulates sequence generation as a \textit{deliberate, step-wise decision-making process}, enabling the exploration of multiple design paths and exploitation of promising candidates through self-evaluation, lookahead, and backtracking. We propose a two-stage \textit{focus-and-grounding} action mechanism that decouples position selection and residue generation. To efficiently evaluate intermediate states, we introduce a \textit{jumpy denoising} strategy that avoids full rollouts. Built upon pretrained protein language models, ProtInvTree supports flexible test-time scaling by expanding the search depth and breadth without retraining.
    Empirically, ProtInvTree outperforms state-of-the-art baselines across multiple benchmarks, generating structurally consistent yet diverse sequences, including those far from the native ground truth.
    
    % It explores multiple different design paths via self-evaluates choices such as looking ahead, backtracking, and evaluation.
    % explores multiple trajectories while exploiting feedback to ensure structural consistency.
    % It introduces three key innovations: (i) a rollout-based formulation of iterative mask-infilling, 
    
    % (ii) a two-stage \textit{focus-and-grounding} action mechanism that decouples position selection and residue generation, 
    % and (iii) a \textit{jumpy denoising} strategy for efficient reward evaluation without full rollout. 
    
\end{abstract}

\section{Introduction}

% 讲inverse folding的定义和问题重定义
% 然后讲目前inverse folding方法 问题都是在比较recovery，然而这个任务本身是一对多的问题；
% 因此，我们重新定义问题的目标是保证结构一致性的同时提升diversity（贡献1
Proteins are 3D folded linear chains of amino acids that perform essential biological functions, such as metabolic control, transmitting signals, and regulating cellular processes~\citep{huang2016coming, bryson1995protein}.
Designing sequences of amino acids that fold into a desired protein structure, also known as protein “inverse folding” (IF)~\citep{yue1992inverse}, is a crucial task with great potential for protein engineering and synthetic biology~\citep{khakzad2023new,zhu2024generative,chu2024sparks}.
% 目前工作怎么做
Recent deep learning approaches typically recover the native sequence conditioned on a target structure through the following three paradigms: \textit{autoregressive generation}, which models sequence dependencies step-by-step~\citep{ingraham2019generative,jinglearning,tan2022generative,hsu2022learning,dauparas2022robust}; \textit{one-shot prediction}, which directly maps structure to sequence in a single forward operation~\citep{gao2023pifold,mao2024de, gao2024uniif}; and \textit{iterative refinement}, which progressively improves an initial design through multiple passes~\citep{zheng2023structure, gao2024kwdesign,gao2022alphadesign,zhubridge}.
Despite of achieving impressive recovery performance, these methods often overlooked the inherently \textit{one-to-many} nature of the problem~\citep{silva2025fast,murphy2012increasing,hamamsy2023protein}, where multiple distinct amino acid sequences are capable of folding into the same protein backbone structure.
As a result, rather than predicting a single native sequence, it is often desirable to generate a diverse set of sequences while preserving structural consistency.

% 第二段讲如何balance diversity和consistency，我们提议借鉴人类的decision making过程，a deliberate, step-wise design process that balances broad exploration with objective-driven evaluation at each step.
This goal, however, reveals an inherent trade-off: 
while \textit{diversity} prefers broad exploration of the sequence space, \textit{structural consistency} strictly requires a feasible subspace that has good local residue compatibility and global foldability.
% This goal reveals a fundamental trade-off: achieving diversity necessitates broad exploration of the sequence space, while maintaining structural consistency requires adherence to a constrained subspace that supports both local residue compatibility and global foldability.
% 现有方法怎么做的，为什么没做到 trade-off，问题在哪里
% fumulation
% To address this challenge, we advocate for a deliberate, step-wise design process that balances broad exploration with objective-driven evaluation at each step.
% To effectively balance this trade-off, 
To address this challenge, we advocate for a deliberate, step-wise design process that progressively explores the solution space of feasible sequences.
% This is inspired by the “System 2” mode when people engage with decisions~\citep{kahneman2002representativeness,stanovich1999rational,sloman1996empirical}, where the reasoning process is slow, deliberate, and conscious. Decisions are made by comparing multiple alternatives and evaluating their potential outcomes before committing.
Inspired by the dual process theory in cognitive science~\citep{kahneman2002representativeness,stanovich1999rational,sloman1996empirical}, where System 1 is characterized by fast, automatic, and heuristic-driven responses, while System 2 involves slow, deliberate, and analytical reasoning, we propose to model the protein design process as 
% Similarly, we suggest protein design to be 
a \textit{deliberate and iterative} decision-making process that should (1) explore multiple alternatives at each design step rather than one single candidate,
(2) dynamically assess and revise each design step through lookahead and backtracking to optimize the
overall
% whole
designed sequence, and
(3) maintain a structured decision history to support effective credit assignment and multi-step planning. 
% To this goal, tree of thought algorithms naturally support these principles by enabling parallel exploration, trajectory-level reasoning, and reward attribution—making them a strong fit for navigating the combination al landscape of sequence design. Tree-of-thought \cite{tot} algorithms naturally support these principles by enabling parallel exploration, trajectory-level reasoning, and reward attribution—making them well-suited for navigating the combinatorial landscape of protein sequence design.

% (1) maintains and explores diverse alternatives for current choices instead of just picking one, (2) evaluates its current status and actively looks ahead or backtracks to achieve globally optimal outcomes, and (3) organizes the reasoning process in a structured manner to provide a clear path history, making it easier to attribute rewards or errors to specific decisions during multi-step planning.

% rethink

\begin{figure*}[tb]
    \centering
	\includegraphics[width=1\textwidth]{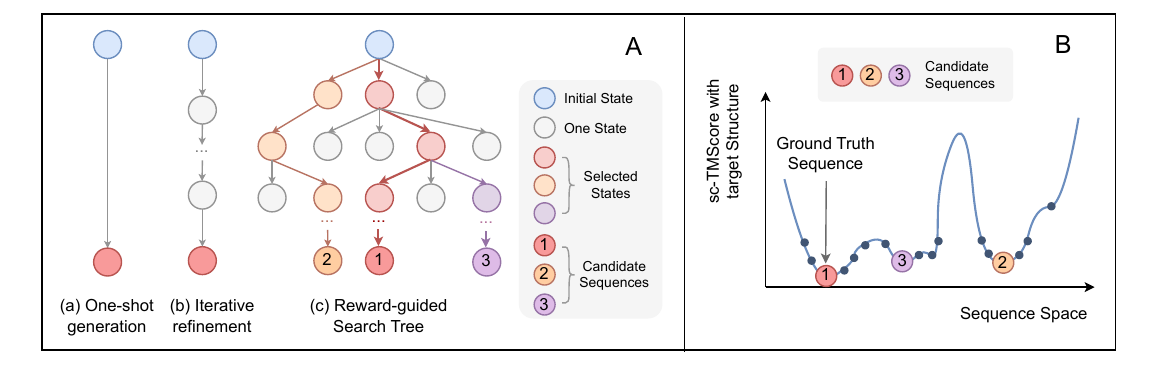}
	\caption{\textbf{(A)} Schematic illustration on various approaches of structure-based protein design. 
    Each node denotes an intermediate sequence conditioned on the target structure, progressively expanding toward full generation. See the complete framework and details in Figures \ref{fig:framework}. \textbf{(B)} Landscape of scTMscore across the sequence space with respect to the target structure. The ground truth sequence is marked, and candidate sequences are distributed across multiple local optima.}
	\label{fig: illus}
    \vspace{-0.45cm}
\end{figure*}
% 这段有点重复
% To enable such a planning-based design process, we revisit the foundations of artificial intelligence and cognitive science, particularly the problem-solving paradigms proposed by Newell, Shaw, and Simon~\citep{newell1959report,simon1971human}. They conceptualized decision-making as a search through a combinatorial problem space, structured as a tree, where each node represents a partial solution and each branch corresponds to a design choice.

To this end, we propose \textbf{ProtInvTree}, a training-free framework for structure-based protein design that formalizes the design process as a sequence of branching decisions and leverages Monte Carlo Tree Search (MCTS) during generation.
Specifically, we  
% leverage a large-scale protein foundation model to 
iteratively perform the design process, sampling multiple decisions at each step and looking ahead to compute reward signals that evaluate the quality of current choices, thereby guiding the overall sequence design.
At each decision step, we introduce a two-stage focus-and-grounding action mechanism that first selects the positions in the sequence to modify (\textit{focus}) and then generates new residues at these positions (\textit{grounding}).
Moreover, we employ fast, jumpy denoising as an evaluation mechanism, efficiently estimating trajectory quality without costly forward model rollouts. 
% Through these designs, ProtInvTree explores the solution space through sampling while guiding generation towards effective subspaces via reward signals.
Through these designs, ProtInvTree is capable of making globally optimal decisions rather than settling for locally optimal ones, which allows it to design novel yet plausible sequences that may deviate significantly from the native sequence (as shown in Fig.~\ref{fig: illus}). 
Additionally, the explicit exploration of design paths may offer potential insights into the interpretability of protein sequences.
To our knowledge, this is the first work to apply a tree-search framework to structure-based protein sequence design.

Empirically, we comprehensively evaluate ProtInvTree across fixed-backbone and de novo protein design tasks.
We demonstrate that ProtInvTree outperforms state-of-the-art baselines and excels in the design of plausible proteins with high structural consistency. 
Besides, it achieves Pareto-optimal trade-offs in both the scTMscore-diversity and scTMscore-novelty.
Notably, we observe that existing approaches  aggressively optimizing for sequence recovery achieve limited novelty at the same sc-TMscore level.
Further analyses reveal that increasing planning depth and expansion width can effectively improve structural consistency, demonstrating that the paradigm of test-time scaling can effectively unlock the potential of pretrained protein language models (PLMs).

% Notably, ProtInvTree supports flexible test-time scaling by adjusting the depth and breadth of the search tree, enabling improved performance without retraining or finetuning.

In summary, our \textbf{contributions} are as follows: 
\begin{itemize}[leftmargin=5.5mm]
    \item We propose \textbf{ProtInvTree}, the first test-time reward-guided tree search framework for protein inverse folding. It formulates protein design as a deliberate, step-wise decision process, enabling \textit{exploration} of multiple trajectories and \textit{exploitation} of promising candidates.
    \item We introduce a two-stage \textbf{focus-and-grounding} mechanism decoupling position selection and residue generation, and a \textbf{fast, jumpy denoising} strategy for efficient reward evaluation.
    \item We demonstrate that ProtInvTree achieves state-of-the-art performance across multiple benchmarks, with a \textbf{test-time scaling capability} that improves both structural consistency and sequence diversity without retraining or finetuning.
    
\end{itemize}

\section{Related Works}

\subsection{Protein Inverse Folding} 
Recently, AI algorithms have spurred a major revolution in modeling protein inverse folding, enabling accurate sequence design conditioned on target structures. 
Existing approaches can be broadly categorized into the following three paradigms based on their generation strategies.
% autoregressive generation, which models sequence dependencies step-by-step; iterative refinement, which progressively improves an initial design through multiple passes; and one-shot prediction, which directly maps structure to sequence in a single forward operation. Each paradigm offers unique trade-offs in terms of accuracy, diversity, and efficiency, shaping the current landscape of structure-conditioned protein design.

\textbf{Autoregressive models} generate sequences residue-by-residue, conditioned on both the 3D structure and previously generated tokens. Pioneering models like GraphTrans~\citep{ingraham2019generative} and GVP~\citep{jinglearning} introduced SE(3)-invariant graph encoders with attention or geometric modules. Later, models such as GCA~\citep{tan2022generative}, ESM-IF~\citep{hsu2022learning}, and ProteinMPNN~\citep{dauparas2022robust} incorporated global context and fine-grained pairwise distance modeling.
% while SurfPro~\citep{songsurfpro} and InvMSAFold~\citep{silva2025fast} focused on surface features and sequence diversity. 
These models offer accurate recovery but suffer from slow inference on long sequences.

\textbf{One-shot models} bypass iterative steps by directly predicting full sequences from structure. PiFold~\citep{gao2023pifold} introduced an efficient graph encoder with an MLP decoder, achieving significant speedups and improved accuracy on long proteins. Uni-IF~\citep{gao2024uniif} generalizes this to multiple molecule types.
% while SurfDesign~\citep{wusurfdesign} enhances recovery through surface-based representations and protein language model integration. 
These models are highly efficient but face challenges in maintaining global structural consistency.

\textbf{Iterative refinement methods} address this by first generating a full sequence and then improving it through multiple steps. AlphaDesign~\citep{gao2022alphadesign} and LMDesign~\citep{zheng2023structure} use confidence-aware predictors and pretrained sequence models for guided refinement. KWDesign~\citep{gao2024kwdesign} combines sequence and structure pretraining with an uncertainty-aware update mechanism. 
Recent works such as BridegIF~\citep{zhubridge} and GraDe-IF~\citep{yi2023graph} apply diffusion to enhance diversity and structural compatibility.
% and RL-DIF~\citep{ektefaie2024reinforcement} 
Fast non-autoregressive diffusion models like PMPnnDiff~\citep{yang2023fast} accelerate inference while preserving accuracy. 
% Despite improved flexibility, these approaches can be computationally demanding due to multiple refinement rounds.

Despite notable advances in protein inverse folding, most efforts focus on training-time improvements, while the inference phase remains underexplored. 
% Techniques like test-time scaling—widely adopted in NLP for enhancing output quality without retraining—have yet to be leveraged in protein design. 
As large protein foundation models emerge, 
% ~\cite{ye2024proteinbench,lin2023,Shin2021}
harnessing test-time computation to boost sequence quality and diversity becomes crucial for incentivizing their full potentials. 
To this end, we propose a novel paradigm based on tree-structured generation, which departs fundamentally from the three existing categories of approaches.

\subsection{Test-time Scaling and MCTS}
Test-time scaling refers to increasing computational resources during inference to enhance model output without modifying its parameters.
This approach has gained significant attention in the field of large language models (LLMs), where performance is improved by generating multiple samples and using reward models for best-solution selection~\citep{snell2024scaling,wu2024empirical,brown2024large}.
Various test-time search methods have been proposed~\citep{ kang2024mindstar,wang2024qimprovingmultistepreasoning}, including random sampling~\citep{selfconsistency}, self-consistency, and tree-search methods~\citep{tot,rap,zhang2024accessing,rstar}.
% like MCTS~\citep{coulom2006efficient}.
Among them, Monte Carlo Tree Search (MCTS), a heuristic search algorithm designed for decision-making tasks~\citep{coulom2006efficient,silver2016mastering},
% schrittwieser2020mastering
has emerged as a powerful technique for structured exploration in the output space of large language models. It enables deliberate reasoning by simulating multiple generation trajectories, selectively expanding promising paths, and integrating reward feedback to guide inference toward high-quality outputs.
Inspired by these advances, we are the first to extend the paradigm of test-time scaling to protein language models (PLMs). Our proposed framework, ProtInvTree, leverages reward-guided tree search to perform deliberate, step-wise protein sequence generation, enabling test-time scaling for improved structural consistency and diversity.
% \clearpage
\section{Preliminaries}

\paragraph{Problem Definition.} The protein inverse folding problem seeks to determine the amino acid sequence $\bx$ that folds into a given target structure $\bc$. Here, $\bx=[x_1,x_2,\dots,x_L]$ represents the sequence of $L$ residues, where $x_i\in\{1,2,\dots,20\}$ denotes the type of the $i$-th residue. The structure $\bc=[c_1,c_2,\dots,c_n]\in\R^{n\times 4 \times 3}$ specifies the Cartesian coordinates of the backbone atoms (N, C-$\alpha$, C, and optionally O) for each residue $\bc_i$. The inverse folding problem can be formally expressed as:
\begin{equation}
    f_{\theta}: \bc \rightarrow \bx,
\end{equation}
where $\theta$ is the learnable parameter. Given that homologous proteins invariably exhibit similar structures, the solution for a given structure is not unique~\citep{hamamsy2023protein}. Hence, an ideal model should be capable of learning the underlying mapping from protein backbone structures to their corresponding sequence distributions $p_{\theta}(\bx \vert \bc)$.
% Previous PiFold \cite{gao2023pifold} and UniIF \cite{gao2024uniif} models address this problem by one-shot sequence generation, achieving a favorable balance between efficiency and accuracy.

\paragraph{Iterative Denoising.}
\label{sec: iter-deno}
Recent advancements in diffusion and iterative refinement models \citep{gao2022alphadesign,zheng2023structure,gao2024kwdesign,zhubridge} have achieved notable success in protein inverse folding. These methods formulate the task as an iterative denoising process that refines the sequence step by step. 
Following this paradigm, we adopt this strategy to progressively construct the sequence.
Formally, starting from an initially corrupted sequence $\mathbf{x}_0$, the model iteratively denoises the sequence into a complete design $\mathbf{x}_T$ through a series of conditional reverse transitions, where $\bx_0$ and $\bx_T$ differ from the definition in diffusion modeling:
\begin{equation}
    p_\theta(\mathbf{x}_T \mid \mathbf{x}_0, \bc) = \prod_{t=1}^{T} p_\theta(\mathbf{x}_{t+1} \mid \mathbf{x}_t, \bc),
\end{equation}
where $\mathbf{x}_t$ represents the intermediate sequence at step $t$, with a subset of amino acids remaining unfilled (e.g., represented by \texttt{[MASK]} tokens), and $\bc$ denotes the target backbone structure. Each reverse step $p_\theta(\mathbf{x}_{t+1} \mid \mathbf{x}_t, \bc)$ refines the current sequence while preserving the structural context.

\section{ProtInvTree: Deliberate Protein Inverse Folding Framework}
In this section, we propose a reward-guided tree-search framework for deliberate protein inverse folding. We first formulate the iterative denoising as a tree-based Markov decision process (MPD), enabling structured exploration over multiple trajectories (Section~\ref{sec:MDP}). Then we describe the Monte Carlo Tree Search (MCTS) procedure employed to identify diverse and high-quality sequences that are consistent with the target backbone structure (Section~\ref{sec:MCTS}). Finally, we introduce two designs of the action and reward components (Sections~\ref{sec:action} and~\ref{sec:score}), which define how the sequence is updated at each step and how intermediate states are evaluated during the search process. We present the detailed algorithm to formalize the entire framework in the appendix.
% ~\ref{app:Algorithms}.

% We propose ProtInvTree, a novel framework for structure-based protein design that enables deliberate, reward-guided exploration of the sequence space through a search tree (as shown in Fig. x). Each node represents an intermediate sequence state, and each branch corresponds to a plausible refinement trajectory.
% A concrete design of ProtInvTree requires answering four core questions:
% \begin{itemize}[leftmargin=5.5mm]
%     % \item \textbf{Step decomposition} (Section~\ref{sec:MDP}): How to decompose the full sequence generation process into multiple decision steps?
%     \item \textbf{Action design} (Section~\ref{sec:action}): How to design refinement actions from each intermediate state?
%     \item \textbf{State scoring} (Section~\ref{sec:score}): How to evaluate each state effectively and efficiently?
%     \item \textbf{Search strategy} (Section~\ref{sec:MCTS}): How to design a tree search algorithm that balances exploration and exploitation?
% \end{itemize}

\subsection{Tree-based MDP Formulation}
\label{sec:MDP}
As described in Section~\ref{sec: iter-deno}, while the step-wise denoising process is effective, it lacks the ability to incorporate intermediate feedback, track uncertainty, and revise previous decisions. 
To overcome these limitations, we reformulate the iterative denoising process as a \emph{tree-based Markov decision process} for structured, feedback-aware generation. 
In this tree structure, each \textbf{node} represents a state \( \bs_t \), each \textbf{branch} corresponds to an action \( \ba_t \), and each node is assigned a \textbf{value} that reflects the reward \(r_t \) at that state.
Specifically, we define the concepts in the tree search framework as follows:
% Specifically, we define the concepts of state, action, reward, and policy model in tree search framework as follows:
\begin{align*}
    \bs_t &\triangleq (\bc, \bx_{t}), &
    % \ba_t &\triangleq \{(i_k, x_{i_k})\}_{k=1}^{K_t}, 
    \ba_t &\triangleq \{(i_k, x_{i_k})\}_{k=1}^{K_t},&
    r_t &\triangleq R(\bs_t, \ba_t),  &
    \pi(\ba_t \mid \bs_t) &\triangleq p_\theta(\bx_{t+1} \mid \bx_t, \bc).  
\end{align*}
Here, the state $\bs_t$ consists of the target backbone structure $\bc$ and a partially generated sequence $\bx_{t}$. The action $\ba_t$ corresponds to the selection of positions in the sequence and modification of new residues, as detailed in Section~\ref{sec:action}. 
The reward $r_t$ is computed by a reward function \( R(\bs_t, \ba_t) \), which evaluates the structural consistency of the modified sequence, as described in Section~\ref{sec:score}.   
The policy model $\pi(\ba_t \mid \bs_t)$ generates the next partial sequence $\bx_{t+1}$ based on the current state $\bs_t$. It is parameterized by a structurally modulated Protein Language Model (PLM).
A trajectory in the multi-step  Markov decision process is defined as a sequence of state-action-reward transitions:
\[
\tau = [(\bs_0, \ba_0, r_0), (\bs_1, \ba_1, r_1), \ldots, (\bs_T, \ba_T, r_T)],
\]
where each transition corresponds to an incremental update of the sequence.
% Our goal is to identify diverse and high-quality trajectories, enabling exploration of multiple sequence candidates that satisfy structural constraints. 
% To this end, we adopt a tree-structured Markov decision process, where each \textbf{node} represents a state $\bs_t$, each \textbf{branch} corresponds to an action $\ba_t$, and each node is assigned a \textbf{value} reflecting the reward $\br_t$ at that state. 
By reformulating the sequence design process from a linear chain into a tree structure, our framework enables deliberate planning over multiple generation trajectories, facilitates the incorporation of intermediate feedback from structural evaluations, and supports systematic revision of prior design decisions.

\subsection{Reward-guided Tree Search}
\label{sec:MCTS}
\begin{figure}[h]
    \centering
    \includegraphics[width=1.0\textwidth]{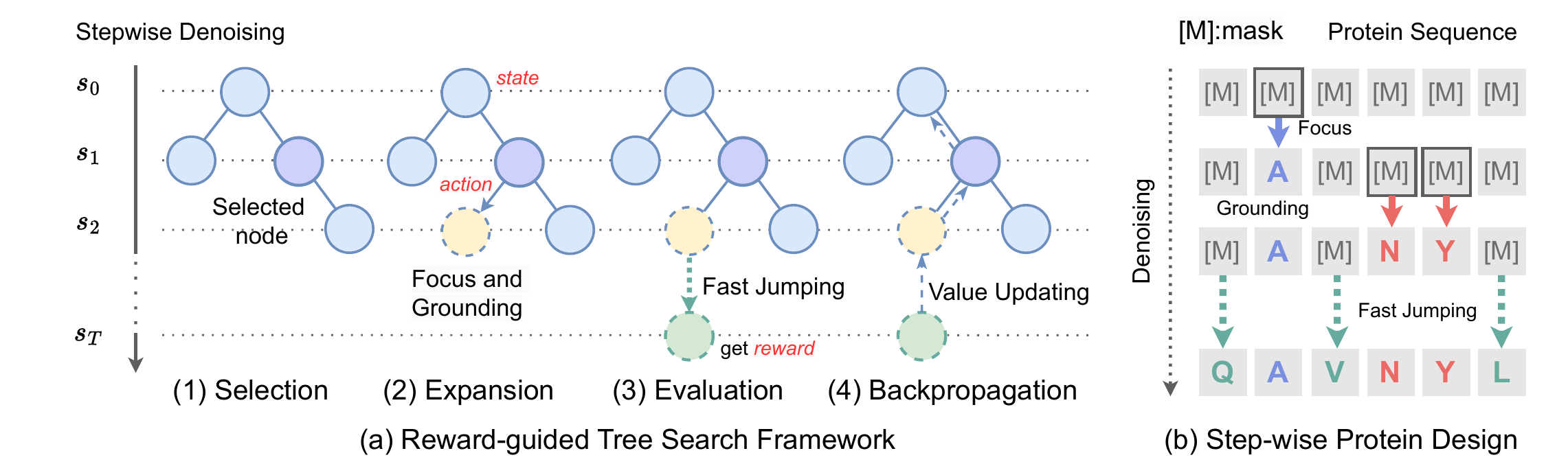}
    \caption{The framework of ProtInvTree. (a) The four steps of reward-guided tree search—\textit{Selection, Expansion, Evalution, and Backpropagation}—are illustrated on a partial denoising tree. Each node corresponds to a partially denoised subsequence. After a new node is expanded, “jumpy” denoising is performed to quickly estimate its value, which is then backpropagated along the path in the tree. (b) Illustration of how a sequence is generated step by step. Masked tokens in the sequence are progressively infilling through a focus-and-grounding mechanism.} 
    \vspace{-1em}
    \label{fig:framework}
\end{figure}

In our approach, the reward-guided tree search process operates as an iterative procedure.
As illustrated in Figure~\ref{fig:framework}, it comprises four key steps: selection, expansion, evaluation, and backpropagation. 
The details of each step are described as below.

\vspace{-0.5em}

\vspace{-0.1em}
\paragraph{Selection.} The selection process begins at the root node $s_0$ and identifies the leaf node with the highest exploration potential, determined by the UCT (Upper Confidence Bounds applied to Trees)~\cite{uct} score. The UCT score is computed as follows:
\begin{equation}
    UCT(\bs_t) = V(\bs_t) + w \sqrt{\frac{\ln N(\bp)}{N(\bs_t)}},
    \label{eq:uct}
\end{equation}
where $w$ is a hyperparameter that balances exploitation (i.e., node value $V(\bs_t)$) and exploration (i.e., visit count $N(\bs_t)$), and $\bp$ denotes the parent node of $\bs_t$.

\vspace{-0.5em}
\paragraph{Expansion.} After selecting the node with the highest UCT score, it is expanded by generating multiple child nodes. 
Conditioned on the current state $\bs_t$, which consists of the target structure $\bc$ and current sequence $\bx_t$, the policy model samples $K$ candidate sequences for the next step: 
% to construct next-step states as follows
\vspace{-0.15em}
\begin{equation}
    \{\bx_{t+1}^{(k)}\}_{k=1}^K \sim \pi_\theta(\ba_{t} \mid \bs_t) \triangleq p_\theta(\bx_{t+1} \mid \bx_t, \bc).
\end{equation}
\vspace{-0.15em}
% These candidate sequences $\{\bx_{t+1}^{(k)}\}_{k=1}^K$ 
% % are then used to construct next-step states, 
% each correspond to a new child state $\bs_{t+1}^{(k)}$ in the tree.
Each candidate sequence $\bx_{t+1}^{(k)}$ constitutes a new child state $\bs_{t+1}^{(k)} = (\bc, \bx_{t+1}^{(k)})$, which is added to the search tree as an expansion of the selected node. The details of the candidate construction process by policy model are described in Section~\ref{sec:action}.

% where $\bs_t = (\bc, \bx_{T-t})$ denotes the current partial sequence conditioned on the target structure $\bc$.
% The policy model generates $K$ candidate partial sequence , each corresponding to a new child state $\bs_{t+1}^{(k)}$ in the tree.
% Compared to previous work~\cite{} mainly focused on a single-step or linear generation paradigm, our approach leverages tree expansion to explore multiple potential generations for fully exploiting the search space.
\vspace{-0.1em}
\paragraph{Evaluation.} Each expanded node is evaluated to determine its value $V(\bs_{t+1})$. As described in Equation~\ref{eq:reward}, we first perform rollouts that complete the state $\bs_{t+1}$ via sampling $m$ fully generated sequences, and then assess them with a reward model, assigning the average reward $r_{t+1}$ as the node value $V(\bs_{t+1})$. The details of the reward function and evaluation process are provided in Section~\ref{sec:score}.

% These sequences, denoted as $\mathcal{C_\mathrm{r}}(s_{t+1})$, are subsequently evaluated by the reward model, which assigns a reward score to each sequence.
% The average score $r$ among $m$ completed sequences is assigned as the initial value $V(s_{t+1})$ of the corresponding child node $s_{t+1}$.
\vspace{-0.1em}
\paragraph{Backpropagation.} After evaluating the expanded nodes, their values are backpropagated along the traversal path to update the visit counts and value scores of the ancestor nodes $\bs_j~(0 \leq j \leq t)$. The updates are performed using the following equations:
\begin{align}
    N_{\text{new}}(\bs_j) &= N_{\text{old}}(\bs_j) + 1, 
    \label{eq:backn}
    \\
    V_{\text{new}}(\bs_j) &= \frac{V_{\text{old}}(\bs_j)N_{\text{old}}(\bs_j) + r_{t+1}}{N_{\text{new}}(\bs_j)},
    \label{eq:backv}
\end{align}
% \begin{align}
%     V_{\text{new}}(\bs_j) &= \frac{V_{\text{old}}(\bs_j)N_{\text{old}}(\bs_j) + r_{t+1}}{N_{\text{new}}(\bs_j)},
%     \label{eq:backv}
% \end{align}
where $N_{\text{old}}(\bs_j)$ and $V_{\text{old}}(\bs_j)$ represent the previous visit count and value score of node $\bs_j$, respectively, and $r_{t+1}$ is the reward obtained during the evaluation step.

The four stages described above are performed iteratively until the terminal state is reached. We define two termination conditions for MCTS as follows: (1) The maximum number of MCTS iterations, $M$, is reached. (2) A terminal node is encountered with a reward exceeding a predefined threshold, indicating strong structural consistency and high-quality design.
Once the tree search is complete, the optimal path is selected greedily by prioritizing nodes with the highest scores.

\subsection{Focus-and-Grounding Action}
\label{sec:action}

To generate candidate transitions from each intermediate state $\bs_t$, we propose a two-stage \textit{Focus-and-Grounding} action mechanism (see illustration in Fig.~\ref{fig:framework}b). At each denoising step, the model explicitly decomposes the sequence updating process into identifying \textbf{where} to modify (Focus) and determining \textbf{what} token to generate at the selected position (Grounding).

Formally, the \textit{Focus} operation $\mathcal{F}(\cdot)$ defines a position selection distribution \( p_\phi(i \mid \bs_t) \) over all positions, from which the top-$K_t$ positions with the highest probabilities are selected:
% \begin{equation}
%     \mathcal{F}(\bs_t) = \operatorname{argsort}_{i \in \{1, \dots, L\}}^{K_t} \left( p_\phi(i \mid \bs_t) \right),
% \end{equation}
% \vspace{-0.001em}
\begin{equation}
    \mathcal{F}(\bs_t) = \operatorname{argsort}_{i \in \{1, \dots, L\}} \left( p_\phi(i \mid \bs_t), K_t \right),
\end{equation}
\vspace{-0.1em}
% \begin{equation}
%     \mathcal{F}(\bs_t) = \text{Top}{K_t}_{i \in \{1, \dots, L\}} \left( p_\phi(i \mid \bs_t) \right),
% \end{equation}
where $L$ denotes the sequence length and $i$ indicates the targeted position for refinement. 
Conditioned on the focused positions \( \{i_1, \dots, i_N\} \), the \textit{Grounding} operation defines a distribution over amino acid types, specifying the generated token:
\begin{equation}
    \mathcal{G}(\bs_t, i)= p_\psi(x_i \mid \bs_t, i), \  i \in \mathcal{F}(\bs_t) ,
\end{equation}
where each \( p_\psi(x_i \mid \bs_t, i) \) predicts the residue \( x_i \in \mathcal{V} \) for position \( i \), and \( \mathcal{V} \) denotes the amino acid vocabulary. 
The overall policy is factorized as the product of the Focus and Grounding distributions:
% \begin{equation}
    % \pi_\theta(\ba_t \mid \bs_t) \triangleq p_\psi(x_i \mid \bs_t, i) \cdot p_\phi(i \mid \bs_t).
    
% \end{equation}
\vspace{-0.8em}
\begin{equation}
\pi_\theta(\ba_t \mid \bs_t) = \prod_{i \in \mathcal{F}(\bs_t)} p_\phi(i \mid \bs_t) \cdot p_\psi(x_i \mid \bs_t, i),
\end{equation}
\vspace{-1em}
% The combined action $\ba_t \in \mathcal{A}$ corresponds to the joint selection of a focus position and its grounded token, represented as:
% \begin{equation}
% \ba_t \triangleq (i, x_i).
% \end{equation}

% The policy $\pi_{\theta}$ is defined as:
% \begin{equation}
%     \pi_{\theta}(\ba_t \mid \bs_t) \triangleq p_{\phi}(i \mid \bs_t) \cdot p_{\psi}(x_i \mid \bs_t, i).
% \end{equation}
In practice, the selected position set \( \{i_1, \dots, i_N\} \) is a random subset of sequence positions (more selection strategies comparison is provided in appendix), and each token \( x_i \) is generated by a structurally modulated Protein Language Model (PLM) conditioned on the backbone structure \( \bc \) and the partial sequence context.
This two-stage action design enables precise localization of modifications, ensuring structural coherence and enhancing search efficiency throughout the generation process.

%建议可以根据plddt等指标选择突变位置，这样更新颖

\subsection{Jumpy Denoising for Fast Reward}
\label{sec:score}
In the MCTS procedure, evaluating a node far from a leaf node is challenging, as the intermediate nodes are not fully expanded. This is typically addressed in one of two ways: employing forward dynamics models to simulate complete trajectories, which is computationally expensive, or approximating node values via bootstrapping methods, which are faster but less accurate. Effectively integrating these evaluation strategies into ProtInvTree remains an open challenge.

% To address this, we introduce a \textit{Jumpy Denoising} strategy $\mathcal{J}(\cdot)$ to accelerate the evaluation process, which is a rapid, single-step DDIM-based sampling process:
% \begin{equation}
%     \tilde{\bx}_0 \sim p(\bx_0 \mid \bx_{t}, \bc).
% \end{equation}
To address this, we introduce a \textit{Jumpy Denoising} strategy to accelerate the evaluation process, which is a rapid, single-step DDIM-based sampling process:
\begin{equation}
    \tilde{\bx}_T \sim \mathcal{J}(\bx_{t+1}, \bc),
    \label{eq:jump}
\end{equation}
where $\mathcal{J}(\cdot)$ approximates the reverse denoising distribution $p(\bx_T \mid \bx_{t+1}, \bc)$. Here, $\bx_{t+1}$ is obtained through action $\ba_t$ at step $t$.
% We define the reward function $\mathcal{R}(\ba_t \mid \bs_{t})$ as the structural consistency feedback for the intermediate generation state $\bs_t$ at step $t$. 
We define the reward function $R(\bs_t, \ba_t)$ as the structural consistency feedback obtained by comparing the folding results from the sampled sequence $\tilde{\bx}_T$ and the input structure $\bc$,  %applying action $\ba_t$ to the state $\bs_t$ at step $t$,
% by first applying action $\ba_t$ to state $\bs_t$ to produce the updated sequence $\bx_{t+1}$, and then evaluating its predicted structure, 
formulated as:
\begin{equation}
    R(\bs_t, \ba_t) = \text{TMScore}(f(\tilde{\bx}_T), \bc),
    \label{eq:reward}
\end{equation}
where $f$ is the protein folding algorithm. %and $\tilde{\bx}_T$ and $\bx_{gt}$ denote the rollout sequence and the ground truth sequence, respectively. 
$\text{TMScore}(\cdot, \cdot)$ is a widely used metric for measuring protein structure similarity. 
This jumpy denoising strategy significantly reduces computational overhead while maintaining a reliable approximation of the final reward.

\section{Experiments}
% In this section, we conduct extensive experiments to answer the following questions: 
\label{sec:experi}

\subsection{Experimental Setup}

\paragraph{Datasets.}
We conduct experiments on both \textbf{CATH v4.2} and \textbf{CATH v4.3}~\cite{orengo1997cath}, where proteins are categorized based on the CATH hierarchical classification of protein structure, to ensure a comprehensive analysis. Following the standard data splitting~\citep{ingraham2019generative,hsu2022learning}, CATH v4.2 dataset consists of 18,024 proteins for training, 608 proteins for validation, and 1,120 proteins for testing; CATH v4.3 dataset consists of 16,153 proteins for training, 1,457 proteins for validation, and 1,797 proteins for testing. We also include a set of de novo proteins collected from the CASP15 competition to provide a more realistic assessment. Following the previous work~\citep{gao2023proteininvbench}, we download the public TS-domains structures from CASP15 which consists of 45 structures, namely \textbf{TS45}.

\begin{table}[t]
   \centering
   \small
   % \vspace{-2.5mm}
   \caption{Structure consistency performance comparison between ProtInvTree and different baseline approaches on the CATH 4.2 dataset. The split of "Short", "Single-chain" and "All" is the same as previous works. The \textbf{best} and \underline{suboptimal} results are labeled with bold and underline.}
   \vspace{+0.3em}
   \label{tab:results_cath4.2}

   \resizebox{\linewidth}{!}{%
   \begin{tabular}{llcccccc}
   \toprule
    \multirow{2}{*}{\bf Models} 
   & \multirow{2}{*}{\bf Trainable/Total} 
   & \multicolumn{3}{c}{\bf scTM-score ($\uparrow$)} 
   & \multicolumn{3}{c}{\bf RMSD ($\downarrow$)} \\
   \cmidrule[0.3pt](lr){3-5} \cmidrule[0.3pt](lr){6-8}
   & \bf Params. & Short   & Single-chain  & All  & Short   & Single-chain   & All \\
   \midrule

StructGNN~\citep{ingraham2019generative} & 1.4M/1.4M & 0.616 & 0.646 & 0.751 & 2.439 & 2.702 & 2.327 \\
GraphTrans~\citep{ingraham2019generative} & 1.5M/1.5M & 0.590 & 0.635 & 0.744 & 2.356 & 2.678 & 2.351 \\
GCA~\citep{tan2023global} & 2.1M/2.1M & 0.606 & 0.646 & 0.755 & 2.430 & 2.596 & 2.226 \\
GVP~\citep{jing2021learning} & 0.9M/0.9M & 0.611 & 0.662 & 0.771 & 2.289 & 2.542 & 2.181 \\
ProteinMPNN~\citep{dauparas2022robust} & 1.9M/1.9M & 0.636 & 0.692 & 0.795 & 2.310 & 2.370 & 2.009 \\
AlphaDesign~\citep{gao2022alphadesign} & 3.6M/3.6M & 0.646 & 0.693 & 0.814 & 2.271 & 2.422 & 1.969 \\
PiFold~\citep{gao2023pifold} & 5.8M/5.8M & 0.655 & 0.700 & 0.842 & 2.203 & 2.355 & 1.723 \\
UniIF~\citep{gao2024uniif} & 5.4M/5.4M & 0.660 & 0.709 & 0.845 & 2.168 & 2.298 & 1.680 \\
LM-Design (ESM-1b)~\citep{zheng2023structure} & 6.9M/650M & 0.663 & 0.714 & 0.849 & 2.150 & 2.240 & 1.638 \\
KW-Design (ESM-2)~\citep{gao2024kwdesign} & 54.49M/650M & \underline{0.676} & \underline{0.729} & \underline{0.858} & \underline{2.101} & \underline{2.148} & \underline{1.566} \\
\cmidrule[0.5pt](lr){1-8}
ESM-3~\citep{hayes2025simulating} & 1.4B/1.4B & 0.668 & 0.692 & 0.816 & 2.060 & 2.387 & 2.135 \\
\textbf{ProtInvTree} (ESM-3) & 0M/1.4B & \textbf{0.768} & \textbf{0.800} & \textbf{0.881} & \textbf{1.902} & \textbf{2.136} & \textbf{1.513} \\
\bottomrule
\end{tabular}
}
\vspace{-0.7em}
\end{table}

% \vspace{-1cm}

\vspace{-0.1cm}
\paragraph{Evaluation Metrics.} For evaluation metrics, we use \textbf{sc-TMscore}~\cite{zhang2005tm} and \textbf{RMSD}~\cite{carugo2003root} to evaluate structural consistency. We define \textbf{diversity} as the average proportion of differing residues across all pairs of generated sequences and define \textbf{novelty} as $ 1 - \textbf{recovery}$.
Details of all metrics are provided in the appendix.
% ~\ref{app:metrics}.
Following previous studies~\citep{ingraham2019generative,hsu2022learning}, we report them on three settings, namely short proteins (length $\leq 100$), single-chain proteins (labeled with 1 chain in CATH), and all proteins.

\begin{table}[t]
\centering
\begin{minipage}[t]{0.485\textwidth}
  \centering
  \caption{Structural consistency comparison between ProtInvTree and baseline approaches on CATH 4.3 datasets. The \textbf{best} and \underline{suboptimal} results are labeled with bold and underline.}
  \vspace{+0.3em}
  \label{tab:results_cath4.3}
  \resizebox{\textwidth}{2cm}{
  \begin{tabular}{lrr}
    \toprule
    \textbf{Model} & \textbf{scTM-score} ($\uparrow$) & \textbf{RMSD} ($\downarrow$) \\
    \midrule
    StructGNN~\citep{ingraham2019generative} & 0.693 & 2.563 \\
    GraphTrans~\citep{ingraham2019generative} & 0.690 & 2.614 \\
    GCA~\citep{tan2023global}  & 0.698 & 2.525 \\
    GVP~\citep{jing2021learning}  & 0.713 & 2.509 \\
    ProteinMPNN~\citep{dauparas2022robust}  & 0.743 & 2.238 \\
    AlphaDesign~\citep{gao2022alphadesign}  & 0.749 & 2.230 \\
    PiFold~\citep{gao2023pifold}  & 0.785 & 1.949 \\
    % UniIF~\citep{gao2024uniif}  & 0.807 & 1.864 \\
    % LM-Design (ESM-1b)~\citep{zheng2023structure}  & 0.809 & 1.812 \\
    KW-Design (ESM-2)~\citep{gao2024kwdesign}  & \underline{0.818} & \underline{1.751} \\
    \midrule
    ESM-3~\citep{hayes2025simulating} & 0.775 & 2.074 \\
    ProtInvTree (ESM-3) & \textbf{0.835} & \textbf{1.702} \\
    \bottomrule
  \end{tabular}}
\end{minipage}
\hfill
\begin{minipage}[t]{0.485\textwidth}
  \centering
  \caption{De novo protein design results on TS45 datasets. We compare structural consistency of the following methods. The \textbf{best} and \underline{suboptimal} results are labeled with bold and underline.}
  \vspace{+0.3em}
  \label{tab:results_casp15}
  \resizebox{\textwidth}{2cm}{
  \begin{tabular}{lrr}
    \toprule
    \textbf{Model} & \textbf{scTM-score} ($\uparrow$) & \textbf{RMSD} ($\downarrow$) \\
    \midrule
    StructGNN~\citep{ingraham2019generative} & 0.631 & 3.336 \\
    GraphTrans~\citep{ingraham2019generative} & 0.618 & 3.276 \\
    GCA~\citep{tan2023global} & 0.660 & 3.226 \\
    GVP~\citep{jing2021learning} & 0.652 & 3.245 \\
    ProteinMPNN~\citep{dauparas2022robust} & 0.668 & 3.142 \\
    AlphaDesign~\citep{gao2022alphadesign} & 0.660 & 3.167 \\
    PiFold~\citep{gao2023pifold} & 0.699 & 2.875 \\
    KWDesign (ESM-2)~\citep{gao2024kwdesign} & \underline{0.711} & \underline{2.643} \\
    \midrule
    ESM-3~\citep{hayes2025simulating} & 0.690 & 2.958 \\
    ProteinInvTree (ESM-3) & \textbf{0.724} & \textbf{2.513} \\
    \bottomrule
  \end{tabular}}

\end{minipage}
\vspace{-0.5em}
\end{table}

\vspace{-0.1cm}
\paragraph{Baselines.}
We compare ProtInvTree with several state-of-the-art baselines, categorized into three groups: (1) autoregressive models, including StructGNN~\citep{ingraham2019generative}, GraphTrans~\citep{ingraham2019generative}, GCA~\citep{tan2023global}, GVP~\citep{jing2021learning}, and ProteinMPNN~\citep{dauparas2022robust}; (2) the one-shot model, PiFold~\citep{gao2023pifold}, UniIF~\citep{gao2024uniif}; (3) iterative models, including  AlphaDesign~\citep{gao2022alphadesign}, LM-Design~\citep{zheng2023structure}, KW-Design~\citep{gao2024kwdesign}.

\vspace{-0.1cm}
\paragraph{Implementation Details.} All experiments are conducted on NVIDIA-A100 GPUs with 80G memory.
We choose ESM-3~\citep{hayes2025simulating} as our policy model because it is the first protein foundation model that directly supports inverse folding without task-specific fine-tuning. 
Building on this capability, we focus on unleashing the potential of large protein language models (PLMs) through \emph{test-time scaling}.
To ensure fast structural feedback for reward computation, we use ESMFold~\citep{lin2022language} to predict the 3D structures of candidate sequences. 
% This greatly improves the efficiency of reward evaluation within the search tree.
For ProtInvTree, we set the maximum number of MCTS iterations $M$ to 50. The selection numbers $K_t$ at each step follow a cosine schedule. In the UCT algorithm, the weight $w$ balancing the exploration and exploitation is set to 0.01. 
% In the ablation study, we vary the number of expanded nodes $K$ and the tree search depth $D$ to investigate their impact on performance.

\subsection{Benchmarking Fixed Backbone Protein Design}
\paragraph{Structural Consistency.}

% Tree Search vs. Linear Denoising.
We benchmark the fixed backbone protein design task in CATH4.2 and CATH4.3 datasets, reporting the scTM-score and RMSD in Tables \ref{tab:results_cath4.2} and \ref{tab:results_cath4.3}.
ProtInvTree demonstrates superior performance over previous methods. We highlight the following: (1) Although iterative refinement models have significantly outperformed previous autoregressive and one-shot baselines, the proposed tree-based generation framework (ProtInvTree) further achieves substantial improvements, demonstrating the effectiveness of branching exploration over linear refinement. (2) ProtInvTree enhances inference based on the frozen ESM-3 model, requiring no additional trainable parameters, yet achieving the strongest performance. Compared to ESM-3 alone, when both methods set the same number of iterative steps, ProtInvTree improves the scTM-score by 18.3\% (short), 17.6\% (single-chain), and 7.8\% (all) in CATH 4.2. (3) While the improvement in scTMscore is expected due to its use as the reward function during search, we further evaluate RMSD as an independent structural metric, which consistently supports the effectiveness of ProtInvTree.

\begin{figure}[tb]
    \centering
    \vspace{-0.5em}
    \includegraphics[width=0.9\textwidth]{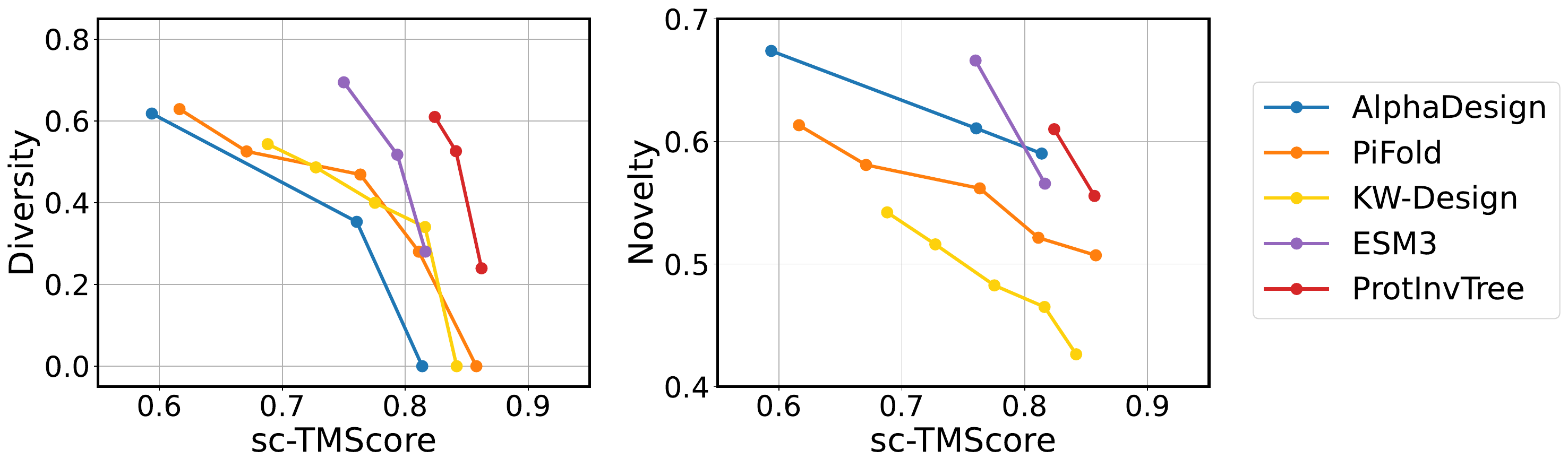}
    \caption{Pareto comparison of structural consistency (sc-TMScore) against diversity (left) and novelty (right) across different protein sequence design methods. Each curve represents a specific method evaluated under different sampling temperatures.}
    \label{fig:pareto}
    \vspace{-1em}
\end{figure}

\begin{figure}[h]
    \centering
    \includegraphics[width=1\textwidth]{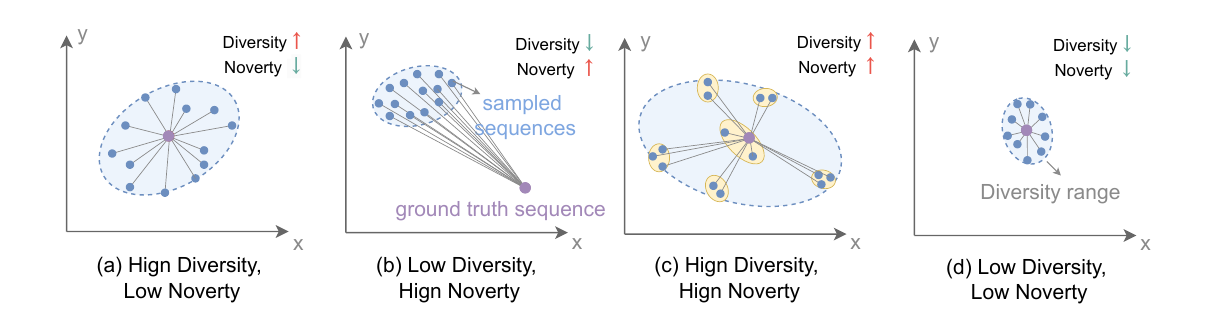}
    \caption{Conceptual illustration of the difference between diversity and novelty of the generated sequences. Each blue dot represents a generated sequence, and the purple dot represents the ground truth sequence. 
    Assuming all generated sequences in this plane share similar structural consistency, 
    The gray circular boundary indicates the diversity range among the generated samples, while the gray lines connecting each sample to the ground truth reflect their novelty.
    % Diversity quantifies variation among generated samples, while novelty reflects deviation from the ground-truth sequence. 
    }
    \label{fig:noverty_diversity}
    \vspace{-0.5em}
\end{figure}

% \vspace{-0.5em}
\paragraph{Balance between Structural Consistency and Diversity \& Novelty.}
Figure~\ref{fig:pareto} illustrates the Pareto frontier between structural consistency (measured by sc-TMScore) and two key sequence-level objectives: \textit{diversity} (left) and \textit{novelty} (right). 
% To ensure a representative and fair comparison across modeling paradigms, we select the best-performing representative from each class of baseline methods: AlphaDesign (autoregressive models), PiFold (one-shot models), and KW-Design (iterative refinement models). 
% We also include ESM-3, the protein foundation model used by our method, to isolate the effectiveness of our inference strategy. 
We highlight our primary findings as follows: (1) ProtInvTree achieves \textbf{Pareto-optimal performance} in both the diversity–scTMscore and novelty–scTMscore spaces, outperforming all baselines across the trade-off frontier. (2) \textbf{Compared to ESM-3}, while improvements in scTMscore are expected due to the reward-guided sampling, we observe that ProtInvTree also achieves significantly higher diversity and novelty even at comparable levels of structural consistency. (3) Notably, when at comparable sc-TMScore levels, the baselines of AlphaDesign, PiFold, and KW-Design exhibit progressively \textit{higher diversity and lower novelty}. This highlights a fundamental distinction between the two metrics: \textbf{diversity} measures variation within the set of generated sequences, whereas \textbf{novelty} reflects deviation from the native (ground-truth) sequence. As shown in Fig.~\ref{fig:noverty_diversity}, baseline methods optimized with a recovery loss tend to converge around local optima near the ground-truth sequence, as illustrated in case (a); by contrast, our method can escape this regime and explore multiple diverse and structurally consistent solutions, including those far from the ground-truth sequence, as shown in case (c).

\vspace{-0.5em}

\subsection{De Novo Proteins Design}

Evaluating models on the TS45 dataset allows us to gain a better understanding of the potential of AI models in designing de novo proteins and reveals that different models exhibit non-trivial differences in generalizability. 
% This dataset serves as a rigorous benchmark to assess both the structural fidelity and generalizability of generated sequences across diverse topologies.
We present the quantitative 
results in 
Table~\ref{tab:results_casp15}, which reveal the following:
(1) \textbf{ProtInvTree} outperforms all baseline methods in terms of both sc-TMscore and RMSD, highlighting its superior ability to maintain structural consistency and achieve accurate geometric reconstruction.
(2) We additionally compare ProtInvTree with ESM-3~\citep{hayes2025simulating} to assess the effectiveness of our overall framework beyond the pretrained language model itself. Despite sharing the same pretrained model, ProtInvTree achieves substantially better results, suggesting that test-time reward-guided planning plays a key role in unlocking the full potential of pretrained protein language models.

% \vspace{-2em}

\subsection{Analysis: Diving Deep into ProtInvTree}

% \begin{wraptable}{r}{0.45\textwidth}
% \vspace{-1.2em}
% \centering
% \captionsetup{font=small}
% \caption{Results of de novo protein design. The \textbf{best} and \underline{suboptimal} results are highlighted.}
% \label{tab:results_casp15}

% \resizebox{0.45\textwidth}{!}{
% \begin{tabular}{lrr}
% \toprule
% \textbf{Model} & sc-TMscore ($\uparrow$) & RMSD ($\downarrow$) \\
% \midrule
% StructGNN~\citep{ingraham2019generative} & 0.631 & 3.336 \\
% GraphTrans~\citep{ingraham2019generative} & 0.618 & 3.276 \\
% GCA~\citep{tan2023global} & 0.660 & 3.226 \\
% GVP~\citep{jing2021learning} & 0.652 & 3.245 \\
% ProteinMPNN~\citep{dauparas2022robust} & 0.668 & 3.142 \\
% AlphaDesign~\citep{gao2022alphadesign} & 0.660 & 3.167 \\
% PiFold~\citep{gao2023pifold} & 0.699 & 2.875 \\
% KWDesign~\citep{gao2024kwdesign} & \underline{0.711} & \underline{2.643} \\
% \midrule
% ESM-3~\citep{hayes2025simulating} & -- & -- \\
% ProteinInvTree & \textbf{--} & \textbf{--} \\
% \bottomrule
% \end{tabular}}
% \vspace{-1em}
% \end{wraptable}

\paragraph{Test-time Scaling Analysis.}
% Planning Depth Analysis
% Expand N Analysis

To understand how test-time computation scales with performance, we investigate the effect of two key planning hyperparameters in our framework: \textit{the number of candidate expansions} and \textit{planning depth}.
We can see that as the expansion number and planning depth increase, the sc-TMscore both gradually improve, although the average time consumed also rises to some extent. 
This highlights that scaling the test-time computation can effectively enhance sequence quality through more deliberate search.
However, as the number of planning depths further increases, the sc-TMscore tends to be saturated, as the search converges to high-confidence regions, the diversity of refinable sequences becomes limited, and further refinements yield diminishing structural gains.
Moreover, the diversity in both settings decreases as the expansion number and planning depth increase, revealing the trade-off between structural consistency and sequence diversity.

% thorough exploration and exploitation by the policy model within the search space. 

% Due to the limitation of inherent knowledge, the policy model cannot benefit a lot from conducting more simulations.

% \vspace{-0.5em}
% \begin{figure}[h]
%     \centering
%     \includegraphics[width=0.85\textwidth]{image/sacling_laws.pdf}
%     \caption{Test-time scaling laws analyses. }
%     \label{fig:scaling}
%     \vspace{-1em}
% \end{figure}
% % \vspace{-2em}

%  \begin{wrapfigure}{r}{0.3\textwidth}
%   \centering
%   \vspace{-0.8em}
%   \includegraphics[width=0.3\textwidth]{image/case.pdf}
%   \caption{Case study.}
%   \vspace{-1em}
%   \label{fig:case}
% \end{wrapfigure}

% \vspace*{-0.1em}
\begin{figure}[h]
  \centering
  \begin{minipage}[t]{0.69\textwidth}
    \centering
    \includegraphics[width=\textwidth]{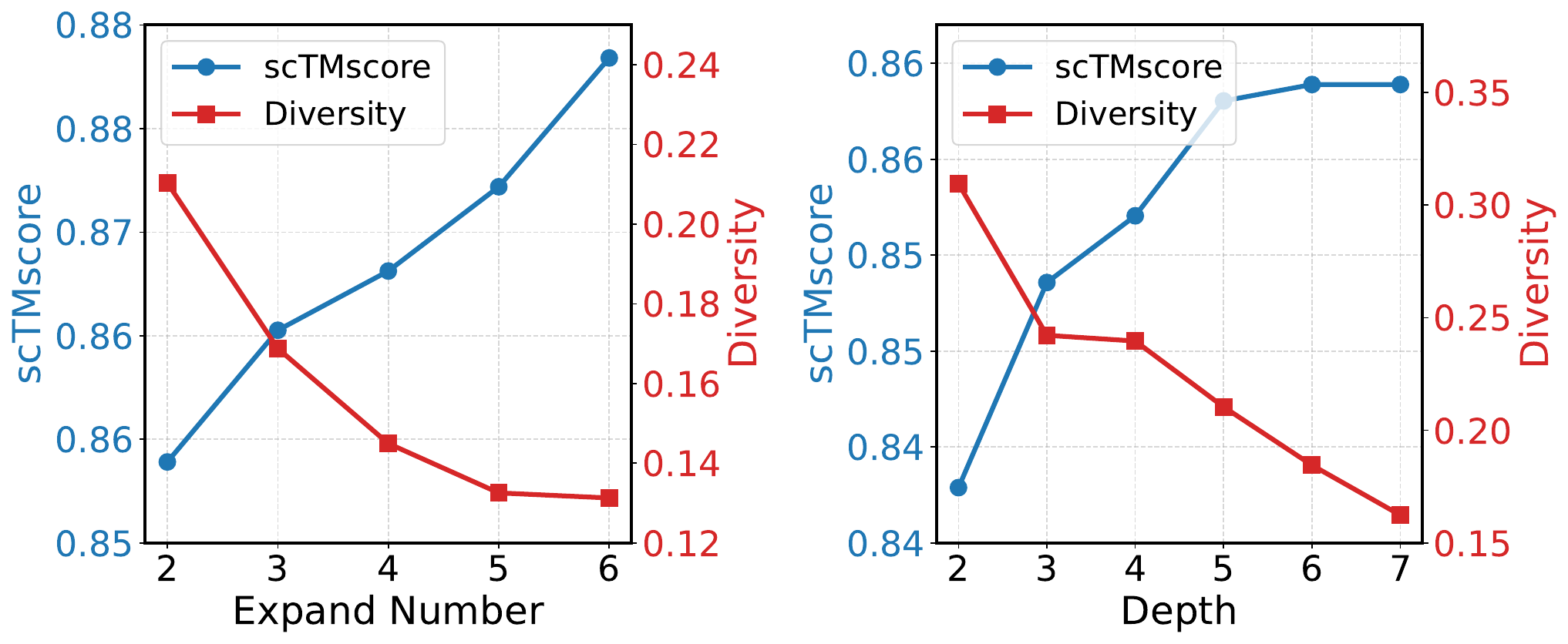}
    \caption{Test-time scaling laws analysis of our ProtInvTree under different expansion numbers (left) and search depths (right).}
    \label{fig:scaling}
  \end{minipage}%
  \hfill
  \begin{minipage}[t]{0.28\textwidth}
    \centering
    \raisebox{0.15em}{\includegraphics[width=\textwidth]{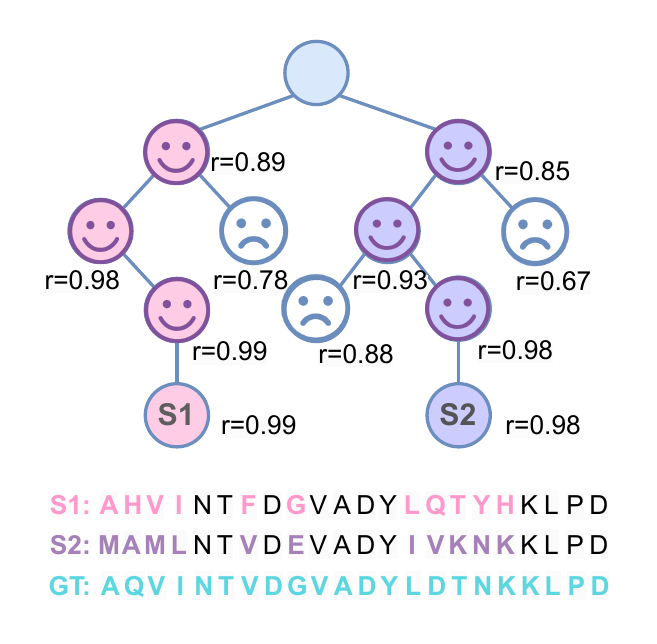}}
    % \captionsetup{font=small}
    \caption{Reward-guided search tree visualization.}
    \label{fig:case_tree}
  \end{minipage}
  \vspace{-0.6em}
\end{figure}

 \begin{wrapfigure}{r}{0.28\textwidth}
  \centering
  \vspace{-1em}
  \includegraphics[width=0.28\textwidth]{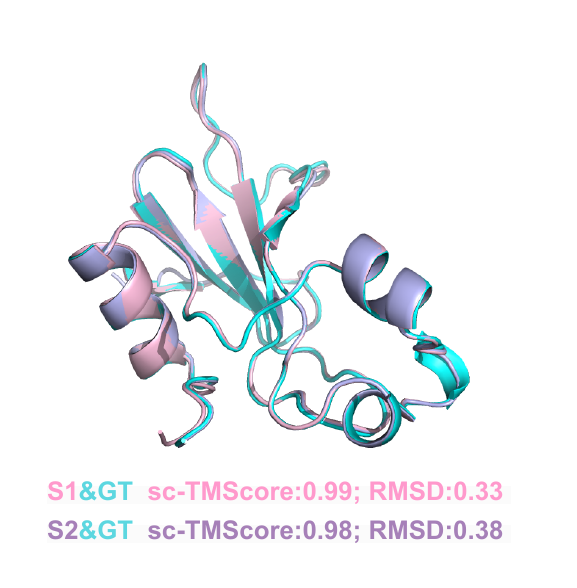}
  \caption{Structural alignment visualization.}
  \vspace{-1em}
  \label{fig:case_structure}
\end{wrapfigure}

\paragraph{Case Study.}
To facilitate understanding of the entire workflow of our proposed ProtInvTree, we visualize a reward-guided search tree in Figure~\ref{fig:case_tree}. Each node represents a partially generated sequence with its predicted reward \(r\), reflecting structural consistency.
The tree showcases how ProtInvTree performs branching exploration guided by reward scores. Two high-reward sequences, \textbf{\textcolor{lightpink}{S1}} and \textbf{\textcolor{lightblue}{S2}}, emerge from different trajectories, with high diversity and novelty, yet achieve high structural rewards (\(r = 0.99\), \(r = 0.98\)). We further compare their predicted 3D structures with the \textbf{\textcolor{cyans}{ground truth}} structure in Figure~\ref{fig:case_structure}.
Its high sc-TMScore and low RMSD demonstrate \textbf{ProtInvTree}'s ability to generate diverse sequence candidates while maintaining structural consistency. This case illustrates how the reward-guided tree search enables efficient exploration of the solution space and selection of structurally faithful, non-trivial designs beyond native recovery.

\section{Conclusion}
% \paragraph{}
% \paragraph{Limitation and Future Work}
We present ProtInvTree, a novel reward-guided tree-search framework for protein inverse folding that explicitly addresses the trade-off between structural consistency and sequence diversity. By reformulating sequence design as a step-wise, decision-making process, ProtInvTree enables the exploration of diverse design trajectories through self-evaluation, lookahead, and backtracking. ProtInvTree shows superior performance across multiple benchmarks, achieving state-of-the-art structural consistency while generating diverse and novel sequences beyond the native ground truth. 
% Future work will focus on extending our framework to a broader range of protein-related tasks beyond fixed-backbone inverse folding. 
% One potential limitation of the proposed ProtInvTree is its current lack of experimental validation in real-world biological settings. We will seek collaborations with experimental laboratories to test the viability and functional relevance of the designed sequences.

% Limitations and future works are shown in Appendix~\ref{app:limitation}.

% \clearpage
\bibliographystyle{plain}  % 或 alpha, unsrt, ieeetr, abbrv 等
\bibliography{refs}        % 这里不加 .bib 后缀

% \section*{References}

% References follow the acknowledgments in the camera-ready paper. Use unnumbered first-level heading for
% the references. Any choice of citation style is acceptable as long as you are
% consistent. It is permissible to reduce the font size to \verb+small+ (9 point)
% when listing the references.
% Note that the Reference section does not count towards the page limit.
% \medskip

% {
% \small

% [1] Alexander, J.A.\ \& Mozer, M.C.\ (1995) Template-based algorithms for
% connectionist rule extraction. In G.\ Tesauro, D.S.\ Touretzky and T.K.\ Leen
% (eds.), {\it Advances in Neural Information Processing Systems 7},
% pp.\ 609--616. Cambridge, MA: MIT Press.

% [2] Bower, J.M.\ \& Beeman, D.\ (1995) {\it The Book of GENESIS: Exploring
%   Realistic Neural Models with the GEneral NEural SImulation System.}  New York:
% TELOS/Springer--Verlag.

% [3] Hasselmo, M.E., Schnell, E.\ \& Barkai, E.\ (1995) Dynamics of learning and
% recall at excitatory recurrent synapses and cholinergic modulation in rat
% hippocampal region CA3. {\it Journal of Neuroscience} {\bf 15}(7):5249-5262.
% }

%%%%%%%%%%%%%%%%%%%%%%%%%%%%%%%%%%%%%%%%%%%%%%%%%%%%%%%%%%%%
\clearpage
\appendix

% \section{Technical Appendices and Supplementary Material}
% Technical appendices with additional results, figures, graphs and proofs may be submitted with the paper submission before the full submission deadline (see above), or as a separate PDF in the ZIP file below before the supplementary material deadline. There is no page limit for the technical appendices.

\section{Algorithms}
\label{app:Algorithms}
The overall workflow of the ProtInvTree is provided in Algorithm~\ref{alg:protinv}.

\begin{algorithm}[h]
\caption{ProtInvTree: Reward-Guided Tree Search for Protein Inverse Folding}
\label{alg:protinv}
 \textbf{Input:} Backbone structure $\bc$, ground truth sequence $\bx_{gt}$, initial sequence $\bx_0$, folding model $f(\cdot)$, PLM policy $\pi_\theta$, reward function $R(\cdot, \cdot)$, max iterations $M$, tree depth $T$, expansion number per node $K$, reward threshold $\tau$\\
 \textbf{Output:} A set of generated sequences $\mathcal{S} = \{\bx^\ast_i\}_{i=1}^{Z}$, where $Z$ denotes the number of generated sequences.

\begin{algorithmic}[1] %[1] enables line numbers
\STATE Initialize root node $\bs_0 = (\bc, \bx_0)$, search tree $\mathcal{T}$ with $\bs_0$, and result set $\mathcal{S} \leftarrow \varnothing $

\FOR{$m = 1$ to $M$}
    \STATE \textcolor{blue}{\textbf{Selection:}} Traverse the tree from $\bs_0$ using UCT to select a promising node $\bs_t$~(Eq.~\ref{eq:uct})
    \FOR{$k = 1$ to $K$}
        \STATE \textcolor{blue}{\textbf{Expansion:}}
        \STATE \quad Sample action $\ba_t^{(k)} \sim \pi_\theta(\ba_t \mid \bs_t)$
        \quad\quad\quad\quad$\triangleright$ Focus-and-Grounding strategies (Sec.~\ref{sec:action})
        \STATE \quad Apply $\ba_t^{(k)}$ to obtain updated sequence $\bx_{t+1}^{(k)}$
        \STATE \quad Construct child state $\bs_{t+1}^{(k)} = (\bc, \bx_{t+1}^{(k)})$
        \STATE \quad Add $\bs_{t+1}^{(k)}$ to tree $\mathcal{T}$ as child of $\bs_t$

        \STATE \textcolor{blue}{\textbf{Evaluation:}}
        \STATE \quad Sample completed sequence: $\tilde{\bx}_T^{(k)} \sim \mathcal{J}(\bx_{t+1}^{(k)}, \bc)$ 
        \quad\quad\quad$\triangleright$ Jumpy denoising (Sec.~\ref{sec:score})
        \STATE \quad Compute reward: $r_{t+1}^{(k)} = \text{TMScore}(f(\tilde{\bx}_T^{(k)}), f(\bx_{gt}))$ 
        \STATE \quad Set node value: $V(\bs_{t+1}^{(k)}) = r_{t+1}^{(k)}$

        \STATE \textcolor{blue}{\textbf{Backpropagation:}}
        \STATE \quad Update visit count $N(\bs_j)$ and value $V(\bs_j)$ (Eq.~\ref{eq:backn})
        \STATE \quad Backpropagate $r_{t+1}^{(k)}$ to update all ancestors of $\bs_{t+1}^{(k)}$ (Eq.~\ref{eq:backv})
        
        \IF{$t+1 = T$ \textbf{and} $r_{t+1}^{(k)} \geq \tau$}
            \STATE Add $\bx_{t+1}^{(k)}$ to result set $\mathcal{S}$
        \ENDIF
    \ENDFOR
\ENDFOR

\STATE \textbf{Return:} Sequence set $\mathcal{S} = \{\bx^\ast_i\}_{i=1}^{Z}$ containing $Z$ high-quality candidates

\end{algorithmic}
\end{algorithm}

\section{Evaluation Metrics}
\label{app:metrics}

In the main paper, we report evaluation results using four metrics: sc-TMscore, RMSD, novelty, and diversity. The descriptions of these metrics are detailed as follows. 
% These metrics collectively measure structural accuracy, geometric fidelity, and sequence-level variation.

\paragraph{sc-TMScore.} 
The structural similarity is the ultimate standard for measuring the quality of the designed sequence. However, the structures of designed protein sequences needed to be predicted by other algorithms, such as AlphaFold~\cite{abramson2024accurate}, RoseTTAFold~\cite{baek2021accurate}, OmegaFold~\cite{wu2022high} and ESMFold~\cite{lin2022language}. The protein folding algorithm itself has a certain inductive bias and will cause some prediction errors, which will affect the evaluation. To overcome the inductive bias, we adapt the self-consistent TM-score (sc-TMscore) metric:
\begin{equation}
    \text{sc-TMScore} = \text{TMScore}(f(\tilde{\bx}), f(\bx)),
    \label{eq:reward}
\end{equation}
where $f$ is the protein folding algorithm and $\text{TMScore}(\cdot, \cdot)$ is a widely used metric~\cite{zhang2005tm} for measuring protein structure similarity. Since the structures of the designed sequence and reference sequence are predicted by the same protein folding algorithm, the model’s inductive bias is expected to be canceled out when calculating the TM-score. This approach results in a more robust metric, called the sc-TMScore, that is less affected by the inductive bias of the protein folding algorithm.

\paragraph{RMSD.}
The standard dissimilarity measure for protein structures is the root mean square deviation (RMSD) of representative atom positions such as $\alpha$-carbons. RMSD is calculated as the square root of the average squared distance between corresponding atoms in two 3D structures:
\begin{equation}
\text{RMSD}(\mathbf{v}, \mathbf{w}) = \sqrt{ \frac{1}{n} \sum_{i=1}^{n} \| v_i - w_i \|^2 },
\label{eq:rmsd}
\end{equation}
where $\mathbf{v} = f(\tilde{\bx})$ and $\mathbf{w} = f(\bx)$ are the predicted 3D structures of the designed sequence $\tilde{\bx}$ and the reference sequence $\bx$, respectively, obtained using a structure prediction algorithm $f$. Here, $v_i$ and $w_i$ denote the 3D coordinates of the $i$-th atom in each structure, and $n$ is the total number of atoms considered (typically backbone or $\alpha$-carbon atoms).
RMSD provides a fine-grained comparison of atomic positions after optimal rigid-body alignment of the two structures. However, it is sensitive to local deviations, such as flexible loops or inaccurate predictions in side-chain packing, and may not fully reflect the overall fold similarity. As a result, RMSD is typically used in conjunction with other metrics such as TM-score to provide a more comprehensive assessment of structural quality.

% \paragraph{RMSD.}
% The standard dissimilarity measure for protein structures is the root mean square deviation (RMSD) of representative atom positions such as $\alpha$-carbons. 
% RMSD is calculated as the square root of the average squared distance between corresponding atoms in two protein structures:
% \begin{equation}
% \text{RMSD}(\mathbf{v}, \mathbf{w}) = \sqrt{ \frac{1}{n} \sum_{i=1}^{n} \| v_i - w_i \|^2 },
% \label{eq:rmsd}
% \end{equation}
% where $v_i$ and ${w}_i$ denote the 3D coordinates of the $i$-th atom in the two structures, respectively, and $n$ is the total number of atoms considered (typically backbone or $\alpha$-carbon atoms).

\paragraph{Novelty.}
We define novelty as the complement of sequence recovery, reflecting the extent to which the generated sequences deviate from the native ground truth:
\[
\text{Novelty} = 1 - \text{Recovery}
\]
where recovery is the fraction of amino acids in the predicted sequence that exactly match the ground-truth sequence at each position, defined as:
\[
\bm{\text{Recovery}} = \frac{1}{n} \sum_{i=1}^{n} \mathbbm{1}(\tilde{x}_i = x_i)
\]

\paragraph{Diversity.}
The average fraction of amino acids that differ between pairs of sequences:

\[
\text{Diversity}(\{ \tilde{\bx}^1, \ldots, \tilde{\bx}^M \}) 
= \frac{2}{NM(M - 1)} \sum_{j=1}^{M} \sum_{k=1}^{j-1} \sum_{i=1}^{N} \mathbbm{1}[\tilde{\bx}^j[i] \ne \tilde{\bx}^k[i]]
\]
\[
= \frac{2}{M(M - 1)} \sum_{j=1}^{M} \sum_{k=1}^{j-1} d_H(\tilde{\bx}^j, \tilde{\bx}^k).
\]

\noindent where $d_H$ is the Hamming distance. We note that sequence diversity alone is not a sufficient measure of an IF method’s quality, as it can be increased arbitrarily at the expense of sample quality (e.g. as measured by structural consistency).

\section{Selection Strategies Comparison}

To analyze the impact of different position selection strategies in the \textit{Focus-and-Grounding} action mechanism, we evaluate several variants for computing the position distribution \( p_\phi(i \mid \bs_t) \), which determines the set of positions \( \{i_1, \dots, i_{K_t}\} \) to be modified at each denoising step.

Specifically, we compare the following approaches:

\begin{itemize}
    \item \textbf{Random sampling:} Positions are selected uniformly at random from the sequence.
    \item \textbf{Autoregressive sampling:} Positions are visited sequentially from left to right in an autoregressive manner.
    \item \textbf{Entropy-based selection:} Positions with the lowest predictive entropy, representing the model’s most confident predictions, are prioritized for update.
\end{itemize}

We integrate each strategy into the \textit{Focus} module \( \mathcal{F}(\bs_t) \), keeping the \textit{Grounding} step unchanged. 
Table~\ref{tab:sampling_comparison} summarizes the quantitative results, showing that all three strategies achieve competitive performance, with \textbf{random sampling} performing surprisingly well despite of its simplicity. This may be because exploring a broader space in the early stages helps avoid premature convergence and encourages greater sequence diversity, which ultimately benefits overall generation quality.

\begin{table}[t]
   \centering
   \small
   \caption{Comparison of different sampling strategies on structure consistency (scTM-score).}
   \vspace{+0.3em}
   \label{tab:sampling_comparison}

   \resizebox{0.5\linewidth}{!}{%
   \begin{tabular}{lccc}
   \toprule
   \bf Sampling Strategy & \bf Random & \bf AR & \bf Entropy \\
   \midrule
   scTM-score ($\uparrow$) & \textbf{0.881} & 0.877 & 0.870 \\
   \bottomrule
   \end{tabular}
   }
   \vspace{-0.7em}
\end{table}

\section{Limitation and Future work}
Future work will focus on extending our framework to a broader range of protein-related tasks beyond fixed-backbone inverse folding. 
One potential limitation of the proposed ProtInvTree is its current lack of experimental validation in real-world biological settings. We will seek collaborations with experimental laboratories to test the viability and functional relevance of the designed sequences.

\section{Broader Impacts}
\label{app:Impacts}
Inverse protein folding models, positioned at the intersection of bioinformatics and computational biology, offer significant potential for advancing both basic research and real-world applications. By enabling the design of protein sequences that reliably fold into desired three-dimensional structures, these models can drive progress across diverse domains. Broader impacts include facilitating structure-based drug discovery, enabling the rational design of enzymes with novel functionalities, and advancing synthetic biology through the creation of custom proteins with tailored properties.

\end{document}